\documentclass[fleqn,10pt]{wlscirep}
\usepackage[utf8]{inputenc}
\usepackage[T1]{fontenc}
\title{Rings of light caused by gravitational waves}

\author[1,*]{Davide Batic}
\author[1,+]{Joud Mojahed Faraji}
\author[2]{Marek Nowakowski}
\author[2,+]{Nicolas Maldonaldo Baracaldo}
\affil[1]{Khalifa University of Science and Technology, Department of Mathematics, Abu Dhabi, 127788, United Arab Emirates}
\affil[2]{Universidad de los Andes, Departament of Physics, Bogota, 111711, Colombia}

\affil[*]{davide.batic@ku.ac.ae}

\affil[+]{these authors contributed equally to this work}


\begin{abstract}
We reconsider the case of the geodesic motion of a massive and
 massless beam of test particles in a gravitational wave. In particular, we use a direct
 Lagrangian approach which simplifies the calculation. Our findings
 differ partly from previously performed calculations  The final
 result can be interpreted as rings of light seen by the observer.
 We give a new interpretation to this picture and show that over large
 distances the effect, albeit rare, could in principle be observable.
\end{abstract}
\begin{document}

\flushbottom
\maketitle
%
%
\thispagestyle{empty}

\section{Introduction}
From the very first moments mankind took interest in astronomy
\cite{history,Lankford}  almost up to today's sophisticated technology of
astrophysical observations, the information about the outer space was
gathered exclusively in the form of the electromagnetic spectrum. It
was, of course, received first in the visible range \cite{visible} and
moved later on to the far infrared \cite{infrared} and high energy
regions \cite{Xray}. Apart from being a breakthrough discovery, the direct
observation of the gravitational wave since 2015 \cite{obgravwave}
marked also a new era of astrophysics known under the name
multi-messenger astrophysics \cite{multimessenger} in which the scientists hope to
study an astrophysical event through the electromagnetic spectrum
enriched with observations of gravitational waves, cosmic rays and
neutrinos.
The gravitational wave physics in future will be directed at tests of
Einstein theory and the precise study of the sources \cite{Ligo}. On
the other hand projects with space based gravitational wave antenna \cite{Lisa}
are progressing giving us the opportunity to probe into inspirals of
extreme mass ratio like a star orbiting a supermassive
black hole. If the expectations come true, our understanding of the
universe will widen considerably. 

Gravitational waves have not only been considered in the standard Einsteinian theory of gravitation \cite{Penrose} but also in the presence of a cosmological constant \cite{MarIv}, strong magnetic fields \cite{Od1} and in the context of modified gravity such as the $f(R)$ family of theories \cite{Od2,Od3}. Effects of gravitational waves on lensing has been studied in \cite{Faraoni1,Faraoni2,Faraoni3}. This is an example of gravitational waves on photons, an effect that will also be treated in the present work. Other effects of gravitational waves in connection with Dark Matter or black holes has been already discussed in \cite{Mota,Yang1,Yang2,Bambi1,Bambi2}.

Albeit being a rather old subject treated in many text books and
specialized literature \cite{books1,books2,books3,books4}, the
gravitational wave physics might still offer some not fully explored
aspects. Here, we have picked up such a challenge and re-considered
the geodesic motion of a massive/massless beam of particles exposed for some time to a gravitational wave \cite{deFelice}.  Our motivation being to
find observational effects, we re-calculated the beam trajectory and
re-interpreted its effects on a screen perpendicular to the propagation direction of the gravitational wave.  We also used a slightly different procedure as compared to the pioneering work of \cite{deFelice}. More precisely, we made use of a direct Lagrangian approach which appears to us more economical and efficient. We do not completely agree with the results presented in
\cite{deFelice} which might be of some importance should the effect discussed in Section IV become observable in the future. In particular, we show that the light defection in a gravitational wave will be seen by an
observer at the distance $R$ as a ring with the radius $RH$ where $H$
is the strain of the gravitational wave.  Over galactic distance scales
such an effect could, in principle, be observable. At the same time we
demonstrate also the limitations of the first order formalism valid
for the weak fields, i.e.,  $g_{\mu \nu}=\eta_{\mu \nu} +h_{\mu
  \nu}$.  Indeed, a more exact formalism is required to calculate
the trajectory of the photon after the gravitational wave has caused
the deflection. 

Finally, we made a small survey of the most relevant textbooks on General Relativity but we could not find an explicit treatment of the geodesic equation in the presence of a gravitational wave as done in \cite{deFelice} and here. It is well known that solving the geodesic equation for a single particle immersed in a weak gravitational wave will not be sufficient to detect effects because it is always possible to construct a transverse traceless chart in which the particle appears stationary at the first order in $h$ \cite{Carroll,Sch}. However, this observation does not apply to our treatment because we consider a beam of particles.

The paper is organized as follows. In section 2, we give our
conventions and notations as well as the equations and solutions governing the geodesic motion together with a brief discussion of the admissible initial data. In section 3, we analyse the case where the light deflected by a gravitational wave displays a ring on an imaginary screen placed at some fixed distance. There, we also give an interpretation of the effect and explain how such an effect could in principle be detected. In section 4, we present our conclusions.

\section{Fundamentals of gravitational wave physics}
This section summarises the basics of the physics and conventions of the gravitational waves which we will need for the discussion of our results. In the following, we adopt  geometrized units, i.e. the speed of light in vacuum, $c$, and the gravitational constant, $G_N$, are set equal to unity. Furthermore, we refer to \cite{Bom,FL} as a general reference for the topic. To be more specific, we consider a gravitational wave represented by a plane wave propagating in the $z$-direction such that the perturbed Minkwoski line element is
\begin{equation}\label{LE}
ds^2= dt^2-\left[1-h_{11}(t,z)\right] dx^2- \left[1+h_{11}(t,z)\right]dy^2+ 2h_{12}(t,z)dx dy- dz^2,\quad h_{12}=h_{21},\quad h_{22}=-h_{11}
\end{equation}
with
\begin{equation}\label{formalt}
h_{\mu\nu}(t,z)=H_{\mu\nu}\sin{\left[k(t-z)+\varphi_{\mu\nu}\right]},
\end{equation}
where $k$, $H_{\mu\nu}$ and $\varphi_{\mu\nu}$ denote the $z$-component of the wave vector, the amplitude of the gravitational wave and its polarization, respectively. The kinematics of a massive or massless particle with four-velocity components $U^\rho=dx^\rho/ds$ in a manifold locally described by the line element (\ref{LE}) can be studied either by means of the geodesic equation 
\begin{equation}\label{geo}
\frac{dU^\rho}{ds}+\Gamma^\rho{}_{\mu\nu} U^\mu U^\nu=0,\quad
\Gamma^\rho{}_{\mu\nu}=\frac{1}{2}\eta^{\eta\lambda}\left(\frac{\partial h_{\nu\lambda}}{\partial x^\mu}+\frac{\partial h_{\mu\lambda}}{\partial x^\nu}-\frac{\partial h_{\mu\nu}}{\partial x^\lambda}\right)+\mathcal{O}(h^2)
\end{equation}
or in terms of the Euler-Lagrange equations
\begin{equation}\label{Lag}
\frac{d}{ds}\left( \frac{\partial\mathcal{L}}{\partial U^\rho} \right)-\frac{\partial\mathcal{L}}{\partial x^\rho}=0 \quad \rho=0,1,2,3,\quad
\mathcal{L}= \frac{1}{2} g_{\mu\nu}U^\mu U^\nu.
\end{equation}
The proof of the equivalence of both sets of equations is trivial and it is known to hold in general. Note that $s$ is the usual proper time when dealing with a massive particle or an affine parameter if a massless particle is considered. By applying Noether's theorem to the Lagrangian (\ref{Lag}), we obtain the following set of equations
\begin{eqnarray}
&&\frac{d{U}^0}{ds}-\frac{(U^1)^2-(U^2)^2}{2}\partial_t h_{11}-U^1 U^2\partial_t h_{12} =0,\label{I}\\
&&\frac{d}{ds}\left[ (1-h_{11})U^1- h_{12} U^2\right] =0,\quad\frac{d}{ds}\left[ (1+h_{11})U^2- h_{12} U^1\right] =0,\label{III}\\
&&\frac{d U^3}{ds}+\frac{(U^1)^2-(U^2)^2}{2}\partial_z h_{11}+U^1 U^2\partial_z h_{12} =0.\label{IV}
\end{eqnarray}
It is noteworthy to observe that the property $\partial_t h_{\mu\nu}=-\partial_z h_{\mu\nu}$ applied to (\ref{IV}) together with the $\rho=0$ geodesic equation leads to the equality $dU^3/ds=dU^0/ds$. A straightforward integration gives $U^3=U^0+C$ with $C$ an integration constant determined by the initial velocities $U^\rho_0$ at $s=0$, that is $C=U^3_0-U^0_0$. Integrating $U^3-U^0=C$ under the conditions $z(0)=0$ and $t(0)=t_0$ leads to the relation
\begin{equation}\label{rltn3}
t-z=t_0+(U^0_0-U^3_0)s,
\end{equation}
which allows to parameterize $h_{\mu\nu}$ in terms of $s$ as follows
\begin{equation}\label{hmn}
h_{\mu\nu}(s)=H_{\mu\nu}\sin{\left[k(t_0+(U_0^0-U_0^3)s)+\varphi_{\mu\nu}\right]},\quad
h_{\mu\nu}(0)=H_{\mu\nu}\sin{\left(kt_0++\varphi_{\mu\nu}\right)}.
\end{equation}
It is a routine computation to verify that the solution to the differential system (\ref{I})-(\ref{IV}) under the condition $U^3_0\neq U^0_0$ is at the first order in $h_{\mu\nu}$
\begin{eqnarray}
U^0&=&U_0^0+\frac{(U_0^1)^2-(U_0^2)^2}{2(U_0^3-U_0^0)}\left[h_{11}(0)-h_{11}(s)\right]+\frac{U_0^1U_0^2}{U_0^3-U_0^0}\left[h_{12}(0)-h_{12}(s)\right],\label{sol1}\\
U^1&=&\left[1-h_{11}(0)+h_{11}(s)\right]U_0^1+\left[h_{12}(s)-h_{12}(0)\right]U_0^2,\quad
U^2=\left[1+h_{11}(0)-h_{11}(s)\right]U_0^2+\left[h_{12}(s)-h_{12}(0)\right]U_0^1,\label{sol3}\\
U^3&=&U^3_0+\frac{(U_0^1)^2-(U_0^2)^2}{2(U_0^3-U_0^0)}\left[h_{11}(0)-h_{11}(s)\right]+\frac{U_0^1U_0^2}{U_0^3-U_0^0}\left[h_{12}(0)-h_{12}(s)\right].\label{sol4}
\end{eqnarray}
At this point a comment is necessary regarding the initial conditions one may consider in (\ref{sol1})-(\ref{sol4}). First of all, the propagation of the gravitational wave does not induce any form of motion to a particle at rest. This is in agreement with the claim made by Flie\ss bach\cite{FL} (see p. $188$ therein) and it can be easily verified by applying the initial data $U^i_0=0$ with $i=1,2,3$ to (\ref{sol1})-(\ref{sol3}). Moreover, the initial condition $U^3_0=U^0_0$ can never be achieved in the massive case and it can only be applied to massless particles. Such a statement requires an explanation. If $U^3_0=U^0_0$, equation (\ref{rltn3}) implies that $t-z=t_0$. This signalizes that $h_{\mu\nu}$ is constant on such planes and hence, $h_{\mu\nu}(0)=h_{\mu\nu}(s)$ there. As a consequence, it follows from (\ref{sol3}) that $U^1=U_0^1$ and $U^2=U_0^2$. If we impose the usual constraint equation
\begin{equation}\label{constraint1}
(\eta_{\mu\nu}+h_{\mu\nu})U^\mu U^\nu=\epsilon,\quad
\epsilon=\left\{
\begin{array}{cc}
0 & \mbox{if}~m=0\\
1 & \mbox{if}~m\neq 0
\end{array}
\right.,
\end{equation}
with $m$ denoting the mass of the particle, then with the help of (\ref{sol3}) and at the first order in the perturbative parameter, we find that
\begin{equation}\label{k1}
\eta_{\mu\nu}U^\mu U^\nu=\epsilon+\left[(U_0^1)^2-(U_0^2)^2\right]\left[h_{11}(0)-h_{11}(s)\right]+2\left[h_{12}(0)-h_{12}(s)\right]U_0^1 U_0^2,
\end{equation}
which evaluated at $s=0$ gives the following constraint for the initial data
\begin{equation}\label{fvv}
\eta_{\mu\nu}U_0^\mu U_0^\nu=\epsilon.
\end{equation}
If $U^3_0=U^0_0$, equation (\ref{fvv}) implies that the combination $(U_0^1)^2+(U_0^2)^2$ can only take on the values $-1$ (massive particle) and $0$ (massless particle). Hence, the aforementioned initial condition, even though it can never be achieved in the massive case, can be imposed to massless particles provided that $U_0^1=U_0^2=0$. Hence, (\ref{I}) and (\ref{IV}) reduce to  $dU^0/ds=0=dU^3/ds$ from which we obtain $U^0=U^0_3$ and $U^3=U^3_0$. This signalizes that for massless particles the geodesic equation admits the constant solution
\begin{equation}\label{konst}
U^0=U^0_0,\quad U^1=0,\quad U^2=0,\quad U^3_0=U_0^0,
\end{equation}
and therefore, a massless particle moving with initial four-velocity vector $(U_0^0,0,0,U_0^0)$ continues to move at constant speed $U^0_0$ in the $z$-direction for any value of the affine parameter. This means that also in this case the propagation of the gravitational wave will not cause any detectable effect on the motion of the particle. We end this section by checking whether a constant solution is allowed for the case of a massive particle. It can be immediately seen that equations (\ref{III}) are satisfied in the case of a constant solution whenever $U_0^1=U_0^2=0$ while (\ref{I}) and (\ref{IV}) reduce to trivial identities. The information about the remaining velocity components can be retrieved from the constraint $\eta_{\mu\nu}U_0^\mu U_0^\nu=1$ and we conclude that the constant solution of the geodesic equation is
\begin{equation}
U^0=\sqrt{1+(U_0^3)^2},\quad U^1=U^2=0,\quad U^3=U_0^3
\end{equation}
in the massive case.

\section{Results}
\begin{figure}[!ht]\label{hicsumego}
\centering
\includegraphics[scale=0.35]{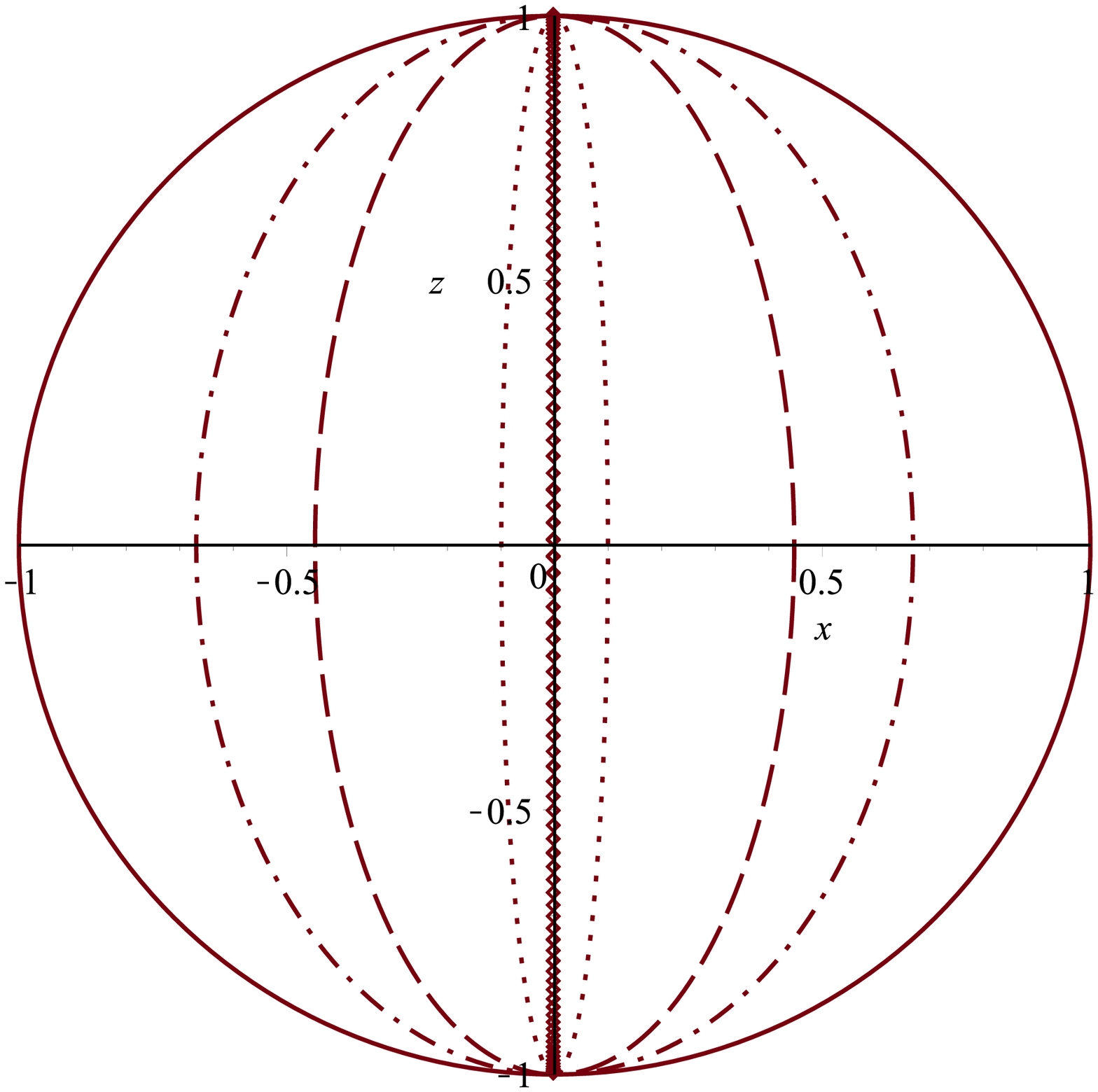}
\includegraphics[scale=0.35]{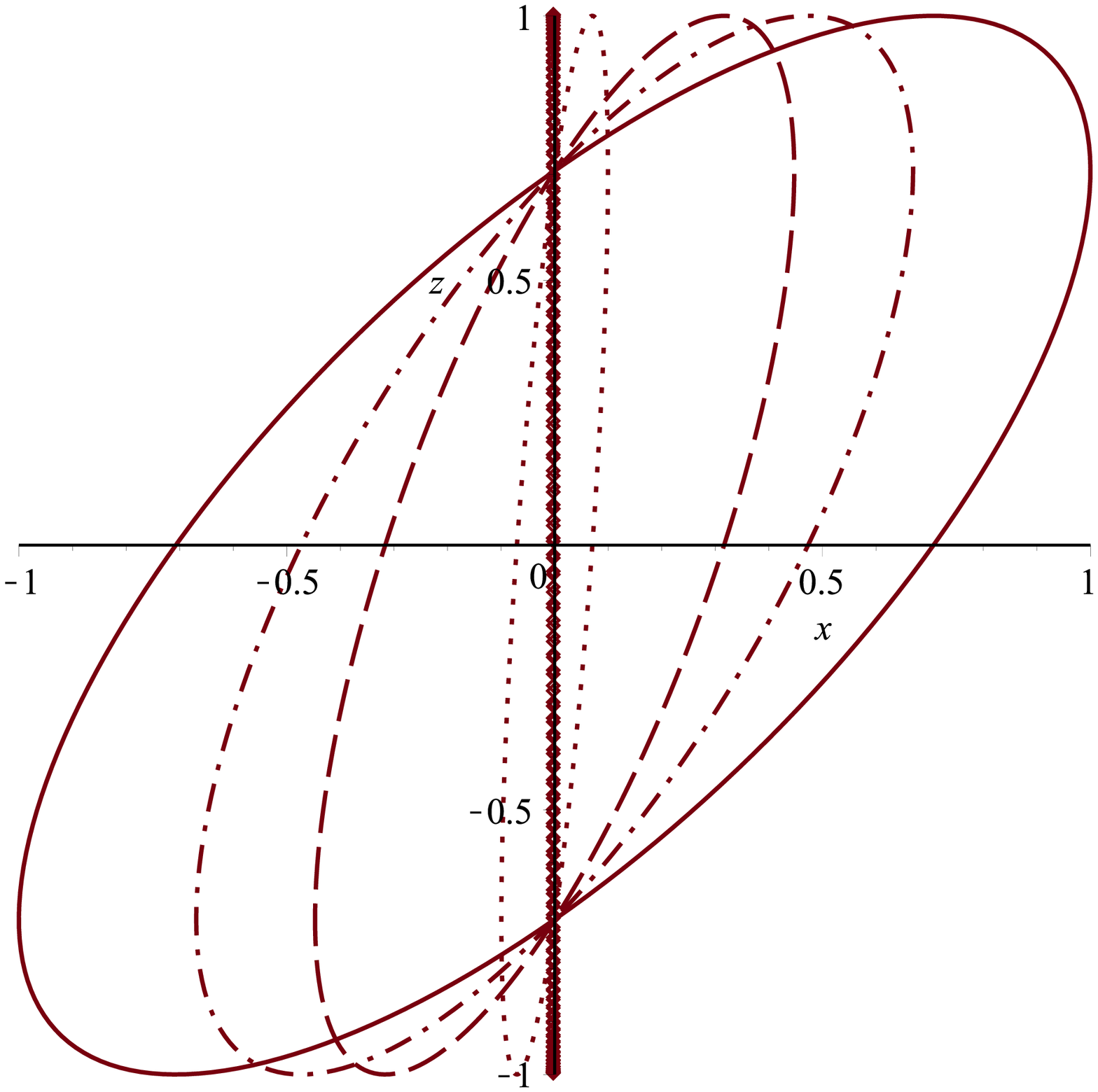}
\caption{\label{figurina}
 Plot of the parametric curves represented by (\ref{par1}) and (\ref{par2}) (left figure) for a gravitational wave (GW) with elliptic polarization $H_{22}=2H_{23}$ and $\varphi_{23}=\varphi_{22}\pm\frac{\pi}{2}$. On the right, we displayed the plot of the parametric curves (\ref{mix1s}) and (\ref{mix2s}) for a GW with $H_{22}=2H_{23}$ and $\varphi_{23}=\varphi_{22}\pm\frac{\pi}{4}$.
  They are traced by the point at which a train of particles successively fired from the origin in the $y$-direction hits a screen positioned perpendicularly to the $y$-axis at $(0,L,0)$. The case of massless particles is represented by the solid circle on the left and the solid ellipse on the right while the massive case is characterized by ellipses with minor axis shrinking towards the $z$-axis as the regime of non-relativistic velocities is approached. For both plots we considered the cases $\gamma=0.9$ (dash-dot ellipse), $\gamma=0.5$ (dash ellipse), $\gamma=0.1$ (dot ellipse) and $\gamma=3.6\times 10^{-5}$ corresponding to Apollo 10 spacecraft (point ellipse) which practically coincides with the line segment $[-1,1]$ on the $z$-axis.
 }
 \end{figure}

In this section, we present new detectable effects of a gravitational wave on massive and massless particles. We start by recalling that one of the pioneer work in this direction was done by de Felice\cite{deFelice} who studied the influence of a gravitational wave on relativistic particles. Here, after discussing some inaccuracies we identified in the aforementioned work, we offer a simple method by which one may amplify a certain effect related to gravitational wave and make it detectable. In order to facilitate the comparison between our findings and the results obtained in de Felice\cite{deFelice}, we observe that the case of a gravitational wave propagating in the $x$-direction can be immediately derived from the already analyzed propagation in the $z$-direction by means of the indices permutation $(0,1,2,3)\longrightarrow(0,2,3,1)$. This leads to the set of equations
\begin{eqnarray}
U^0&=&U_0^0-\frac{(U^2_0)^2-(U^3_0)^2}{2(U_0^0-U_0^1)}h_{22}(0)-\frac{U_0^2 U_0^3}{U_0^0-U_0^1}h_{23}(0)+\frac{(U^2_0)^2-(U^3_0)^2}{2(U_0^0-U_0^1)}h_{22}(s)+\frac{U_0^2 U_0^3}{U_0^0-U_0^1}h_{23}(s),\label{uu1}\\
U^1&=&U_0^1-\frac{(U^2_0)^2-(U^3_0)^2}{2(U_0^0-U_0^1)}h_{22}(0)-\frac{U_0^2 U_0^3}{U_0^0-U_0^1}h_{23}(0)+\frac{(U^2_0)^2-(U^3_0)^2}{2(U_0^0-U_0^1)}h_{22}(s)+\frac{U_0^2 U_0^3}{U_0^0-U_0^1}h_{23}(s),\label{uu2}\\
U^2&=&U_0^2-U_0^2h_{22}(0)-U_0^3h_{23}(0)+U_0^2 h_{22}(s)+U_0^3 h_{23}(s),\label{uu3}\\
U^3&=&U_0^3+U_0^3 h_{22}(0)-U_0^2h_{23}(0)-U_0^3 h_{22}(s)+U_0^2 h_{23}(s)\label{uu4}
\end{eqnarray}
together with the constraint equations (\ref{constraint1}) and (\ref{fvv}). For the following analysis, we underline that the difference $U^0-U^1$ is constant in both the massive and massless cases. More precisely, we have
\begin{equation}\label{wichtig}
U^0-U^1=U^0_0-U^1_0.
\end{equation}
in the massive case, \cite{deFelice} gives the components of the four-velocity vector but they do not appear in linearized form. After linearization they read
\begin{eqnarray}
u^0&=&\frac{1+\alpha^2+\beta^2+E^2}{2E}+\frac{\alpha^2-3\beta^2}{2E}h_{22}(s)+\frac{\alpha\beta}{E}h_{23}(s),\label{ddf1}\\
u^x&=&\frac{1+\alpha^2+\beta^2-E^2}{2E}+\frac{\alpha^2-3\beta^2}{2E}h_{22}(s)+\frac{\alpha\beta}{E}h_{23}(s),\label{ddf2}\\
u^y&=&\alpha+\alpha h_{22}(s)+\beta h_{23}(s),\label{ddf3}\\
u^z&=&\beta-\beta h_{22}(s)+\beta h_{23}(s).\label{ddf4}
\end{eqnarray}
At this point a couple of remarks are in order. First of all, by means of (\ref{ddf1}) and (\ref{ddf2}) we observe that the difference $u^0-u^x$ is correctly predicted to remain constant however the coefficients going together with $h_{22}(s)$ in the first two equations above do not match the corresponding coefficients appearing in (\ref{uu1}) and (\ref{uu2}). The discrepancy can be traced back to a typo in the expression for the auxiliary function $f$ given by $(5)$ in \cite{deFelice} which should be redefined as follows 
\begin{equation}
f=-(1-h_{22})(u^y)^2-(1+h_{22})(u^z)^2+2h_{23}u^y u^z.
\end{equation}
According to the modification above, the constants $E$, $\alpha$ and $\beta$ are connected to the initial four-velocity of the particle by means of the relations
\begin{equation}
E=U^0_0-U^1_0,\quad\alpha=U_0^2,\quad\beta=U_0^3.
\end{equation}
and the numerator $\alpha^2-3\beta^2$ in (\ref{ddf1}) and (\ref{ddf2}) becomes $\alpha^2-\beta^2$ and thus, in agreement with (\ref{uu1}) and (\ref{uu2}). From the set of equations (\ref{ddf1})-(\ref{ddf4}), it can be evinced that \cite{deFelice} assumes $h_{22}(0)=0=h_{23}(0)$ when integrating the geodesic equation while in Section $4$, \cite{deFelice} integrates the corrected version of the aforementioned equations letting $h_{22}(0)\neq 0\neq h_{23}(0)$ in order to derive the trajectory of a massive particle emitted with initial four-velocity vector $u^\mu_0=(u^0_0,0,u^y_0,0)$ from the origin of a Cartesian system when the proper time $\tau=0$! The same mathematical inconsistency emerges from the four-velocity vector in the massless case provided by equation $(6)$ in \cite{deFelice} which are not given there in linearized form. After linearization we find that 
\begin{eqnarray}
k^0&=&\frac{\widetilde{E}^2+\widetilde{\alpha}+\widetilde{\beta}}{2\widetilde{E}}+\frac{\widetilde{\alpha}+\widetilde{\beta}-2\widetilde{\alpha}\widetilde{\beta}}{2\widetilde{E}}h_{23}(s),\\
k^x&=&\frac{\widetilde{E}^2-\left(\widetilde{\alpha}+\widetilde{\beta}\right)}{2\widetilde{E}}+\frac{2\widetilde{\alpha}\widetilde{\beta}-\left(\widetilde{\alpha}+\widetilde{\beta}\right)}{2\widetilde{E}}h_{23}(s),\\
k^y&=&\widetilde{\alpha}+\widetilde{\alpha}h_{22}(s)+\widetilde{\beta}h_{23}(s),\\
k^z&=&\widetilde{\beta}-\widetilde{\beta}h_{22}(s)+\widetilde{\alpha}h_{23}(s).
\end{eqnarray}
The first problem we observe is that already before the linearization occurrs, the difference $k^0-k^x$ does not remain constant. This is is a minor issue  which is easily solved if the original expression for $k^x$ in \cite{deFelice} is replaced by $-k^x$. Two more serious issues are the introduction of the assumption  $h_{22}(0)=0=h_{23}(0)$ without any apparent reason and the independence of the expressions for $k^0$ and $k^x$ on $h_{22}(s)$ while our corresponding results for the components $U^0$ and $U^1$ (see (\ref{uu1}) and (\ref{uu2}) clearly show a dependence on both $h_{22}$ and $h_{23}$. Hence, the formulae provided in $(6)$ by \cite{deFelice} should be taken with extreme caution. This long remark shows that the approach followed by \cite{deFelice} is not mathematically consistent. For this reason, we will first verify that the requirement $h_{22}(0)=0=h_{23}(0)$ cannot be justified on physical grounds and after that, we will indicate how the results obtained in Section $4$ by \cite{deFelice} should be modified. In addition, we also study the trajectory of a beam of massless particle with initial four-velocity vector $u^\mu_0=(u^0_0,0,u^y_0,0)$ directed against a screen orthogonal to the $y$-axis and the shape and size of the bright spot produced on the screen. The same analysis will be performed for beams in the $z$- and $x$-directions and the results compared with the case of propagation along the $x$-axis.\\
As in \cite{deFelice} we consider the motion of a beam of massive particles initially in motion in the $y$-direction with initial four-vector velocity $U_0^\mu=(U_0^0,0,U_0^2,0)$. In this case the parameter $s$ becomes the usual proper time $\tau$. We further suppose that when $\tau=0$ the first particle starts at $x=y=z=0$ and $t(0)=t_0$.  The constraint equation for the massive case gives $U_0^0=\sqrt{1+(U_0^2)^2}$ and from (\ref{wichtig}) we obtain
\begin{equation}
U^0-U^1=\sqrt{1+(U_0^2)^2}
\end{equation}
which integrated gives
\begin{equation}\label{tx}
t-x=t_0+\sqrt{1+(U_0^2)^2}s.
\end{equation}
Taking into account that in the present case $h_{\mu\nu}(t,x)=H_{\mu\nu}\sin{\left[k\left(t-x\right)+\varphi_{\mu\nu}\right]}$, it follows with the help of (\ref{tx}) that
\begin{equation}
h_{\mu\nu}(s)=H_{\mu\nu}\sin{\left[k\left(t_0+\sqrt{1+(U_0^2)^2}s\right)+\varphi_{\mu\nu}\right]}.
\end{equation}
We immediately see that $h_{\mu\nu}(0)=0$ whenever $\varphi_{\mu\nu}=n\pi-kt_0$ with $n\in\mathbb{Z}$. This would imply that the initial time at which an external observer decides to shoot the first particle in the $y$-direction is completely fixed by the polarization of the gravitational wave or the other way around, that is shooting the particle at a certain time $t_0$ influences at least one property of the incoming gravitational wave. Such an argument shows that is not possible to fix $h_{\mu\nu}(0)=0$ as it was done in \cite{deFelice}.\\
Let us consider a massive particles initially moving in the $y$-direction with four-velocity vector $U_0^\mu=(\sqrt{1+(U_0^2)^2},0,U_0^2,0)$. Equations (\ref{uu1})-(\ref{uu4}) become
\begin{eqnarray}
U^0&=&\sqrt{1+(U_0^2)^2}-\frac{(U^2_0)^2}{2\sqrt{1+(U_0^2)^2}}h_{22}(0)+\frac{(U^2_0)^2}{2\sqrt{1+(U_0^2)^2}}h_{22}(\tau),\label{uuu1}\\
U^1&=&-\frac{(U^2_0)^2}{2\sqrt{1+(U_0^2)^2}}h_{22}(0)+\frac{(U^2_0)^2}{2\sqrt{1+(U_0^2)^2}}h_{22}(\tau),\label{uuu2}\\
U^2&=&U_0^2-U_0^2h_{22}(0)+U_0^2 h_{22}(\tau),\label{uuu3}\\
U^3&=&-U_0^2h_{23}(0)+U_0^2 h_{23}(\tau)\label{uuu4}
\end{eqnarray}
where $\tau$ denotes the proper time. The above equations should replace the equations appearing in $(20)$ in \cite{deFelice} where both $h_{22}(0)$ and $h_{23}(0)$ have been assumed to vanish. Suppose that the particle is shot in the $y$-direction from the point $x=y=z=0$ so that $x(0)=y(0)=z(0)=0$ when $\tau=0$ and imagine a screen is placed at $y=L$ from the origin of a Cartesian coordinate system and perpendicularly to the $y$-axis. In the absence of a gravitational wave, the particle would hit the screen at the point $(0,L,0)$. If we integrate (\ref{uuu1}) and restore units, we find the following relation linking the time measured by an external observer with the proper time of the particle
\begin{equation}\label{taut}
t(\tau)=t_0+\frac{2+\gamma^2\left[2-h_{22}(0)\right]}{2\sqrt{1+\gamma^2}}\tau
-\frac{\gamma^2 H_{22}}{2kc(1+\gamma^2)}\left\{\cos{{\left[kc(t_0+\sqrt{1+\gamma^2}\tau)+\varphi_{22}\right]}}-\cos{(kct_0+\varphi_{22})}\right\}
\end{equation}
with $\gamma=U_0^2/c^2$. Finally, integrating (\ref{uuu2})-(\ref{uuu3}) yields
\begin{eqnarray}
x(\tau)&=&-\frac{c\gamma^2 h_{22}(0)}{2\sqrt{1+\gamma^2}}\tau-\frac{\gamma^2 H_{22}}{2k(1+\gamma^2)}\left\{\cos{{\left[kc(t_0+\sqrt{1+\gamma^2}\tau)+\varphi_{22}\right]}}-\cos{(kct_0+\varphi_{22})}\right\},\label{xtau}\\
y(\tau)&=&c\gamma\left[1-h_{22}(0)\right]\tau-\frac{\gamma H_{22}}{k\sqrt{1+\gamma^2}}\left\{\cos{{\left[kc(t_0+\sqrt{1+\gamma^2}\tau)+\varphi_{22}\right]}}-\cos{(kct_0+\varphi_{22})}\right\},\label{ytau}\\
z(\tau)&=&-c\gamma h_{23}(0)\tau-\frac{\gamma H_{23}}{k\sqrt{1+\gamma^2}}\left\{\cos{{\left[kc(t_0+\sqrt{1+\gamma^2}\tau)+\varphi_{23}\right]}}-\cos{(kct_0+\varphi_{23})}\right\},\label{ztau}
\end{eqnarray}
while the solution of the same problem but for a massless particle initially moving in the $y$-direction with four-velocity vector $U_0^\mu=(c,0,c,0)$ from the point $x=y=z=0$ when $s=0$ leads to the following trajectory 
\begin{eqnarray}
x(s)&=&-\frac{c}{2}h_{22}(0)s-\frac{H_{22}}{2k}\left\{\cos{{\left[kc(t_0+s)+\varphi_{22}\right]}}-\cos{(kct_0+\varphi_{22})}\right\},\label{xms}\\
y(s)&=&c\left[1-h_{22}(0)\right]s-\frac{H_{22}}{k}\left\{\cos{{\left[kc(t_0+s)+\varphi_{22}\right]}}-\cos{(kct_0+\varphi_{22})}\right\},\label{yms}\\
z(s)&=&-ch_{23}(0)s-\frac{H_{23}}{k}\left\{\cos{{\left[kc(t_0+s)+\varphi_{23}\right]}}-\cos{(kct_0+\varphi_{23})}\right\}.\label{zms}
\end{eqnarray}
We can follow two different approaches in order to describe how the intersection point between the particle trajectory and the screen will move on the latter as a train of particles is beamed from the origin towards the screen at different emission times $t_0$. Note that $t_0$ enters in the expressions above also through the terms $h_{22}(0)$ and $h_{23}(0)$. For large values of the proper time $\tau\gg 1$ and similarly, for large values of the affine parameter, i.e. $s\gg 1$, equations (\ref{xtau})-(\ref{ztau}) and (\ref{xms})-(\ref{zms}) can be approximated as follows
\begin{eqnarray}
x(\tau)&=&-\frac{c\gamma^2 h_{22}(0)}{2\sqrt{1+\gamma^2}}\tau,\quad~~~x(s)=-\frac{c}{2}h_{22}(0)s,\label{mix1}\\
y(\tau)&=&c\gamma\left[1-h_{22}(0)\right]\tau,\quad y(s)=c\left[1-h_{22}(0)\right]s,\label{mix2}\\
z(\tau)&=&-c\gamma h_{23}(0)\tau,\quad~~~~~~z(s)=-ch_{23}(0)s.\label{mix3}
\end{eqnarray}
Let $\tau_s$ and $\sigma_s$ denote the particular value of the proper time and of the affine parameter at which a massive and a massless particle hit the screen at $y=L$. Simple algebra followed by linearization in the perturbation $h_{\mu\nu}$ gives
\begin{equation}\label{zeit}
\tau_s=\frac{L}{c\gamma}\left[1+h_{22}(0)\right],\quad\sigma_s=\frac{L}{c}\left[1+h_{22}(0)\right].
\end{equation}
From the formulae above, we realize that our initial assumptions $\tau\gg 1$ and $s\gg 1$ will hold, if we require that $\tau_s,\sigma_s\gg 1$. The latter can be always achieved by ensuring that $L\gg 1$ in the appropriate sense. If we substitute (\ref{zeit}) into (\ref{mix1})-(\ref{mix3}) and linearize, we find that the coordinates of the spot on the screen are given by
\begin{eqnarray}
x(\tau_s)&=&-\frac{L\gamma H_{22}}{2\sqrt{1+\gamma^2}}\sin{(kct_0+\varphi_{22})},\quad~~~x(\sigma_s)=-\frac{LH_{22}}{2}\sin{(kct_0+\varphi_{22})},\label{mix1s}\\
y(\tau_s)&=&L,\quad~~~~~~~~~~~~~~~~~~~~~~~~~~~~~~~~~~~~~~y(\sigma_s)=L,\label{mix2s}\\
z(\tau_s)&=&-L H_{23}\sin{(kct_0+\varphi_{23})},\quad~~~~~~~~~~z(\sigma_s)=-L H_{23}\sin{(kct_0+\varphi_{23})}.\label{mix3s}
\end{eqnarray}
At this point a comment is in order. We observe that (\ref{mix3s}) signalizes an interesting feature, namely the displacement of the original spot in the $z$-direction on the screen is independent on whether the impinging particle is massive or massless. This is instead not the case for the displacement in the $x$-direction where we see a sharp distinction between the massive and massless cases. In particular, the effect is larger for the massless case as it can be seen from the following relation
\begin{equation}
x(\sigma_s)=\sqrt{1+\frac{1}{\gamma^2}}~x(\tau_s)
\end{equation}
obtained from (\ref{mix1s}). In order to represent the curve traced on the screen by successively fired particles ($t_0$ becomes a running variable), we will consider a gravitational wave with elliptical polarization. More precisely, we choose $H_{22}=2H_{23}$ and $\varphi_{23}=\varphi_{22}\pm\frac{\pi}{2}$. After rescaling equations (\ref{mix1s}) and (\ref{mix3s}) and letting $\alpha=kct_0+\varphi_{22}$ with $\alpha\in[0,2\pi]$, we end up with the following parametric representations of an ellipse (massive case) and a circle (massless case)
\begin{eqnarray}
x_s&:=&\frac{x(\tau_s)}{LH_{23}}=-\frac{\gamma}{\sqrt{1+\gamma^2}}\sin{\alpha},\quad~~~\widetilde{x}_s:=\frac{x(\sigma_s)}{LH_{23}}=-\sin{\alpha},\label{par1}\\
z_s&:=&\frac{z(\tau_s)}{LH_{23}}=\mp\cos{\alpha},\quad~~~~~~~~~~~~~~~~\widetilde{z}_s=z_s.\label{par2}
\end{eqnarray}
We plotted (\ref{par1}) and (\ref{par2}) for different choices of $\gamma$ in Figure~\ref{figurina}.
 We underline the fact that for massless particles (such has light) continuously fired from some astrophysical source the corresponding spot on the screen will describe a circle of radius $R=LH_{23}$ in the case of an elliptically polarized gravitational wave. As shown in Table~\ref{tableEins}, this effect can be significantly magnified if we imagine the screen to be positioned for example on the surface of the moon while considering light rays coming from distant bright astrophysical objects such as stars and galaxies. Such a magnification is strikingly noticeable if we recall that in LIGO gravitational waves associated to astrophysical sources distant tens of millions of light years from us induce a distorsion of the 4 Km mirror spacing by about $10^{-18}$ m.
 \begin{table}[ht]
\caption{Numerical values of the quantity $LH_{23}$ for a massless particle in the case of an elliptically polarized gravitational wave and for different astrophysical bright objects. The two numbers appearing in the column for $LH_{23}$  refer to choices of a strain $h$ with magnitude $10^{-20}$ \cite{deFelice} and $10^{-22}$ \cite{GWLIGO}, respectively. The distance $L$ has been taken from \cite{Betel,Rigel,Deneb,Vau}. We recall that one megalight-year (Mly) is $10^6$ ly and $1$ ly~$=9.4607\cdot 10^{15}$~m.}
\begin{center}
\begin{tabular}{ | l | l | l | l|}
\hline
$\mbox{Astrophysical source}$ & $L$  & $LH_{23}$ (m) \\ \hline
\mbox{Betelgeuse}  & $548$ ly & $5.18\cdot(10^{-4}\div 10^{-2})$  \\ \hline
\mbox{Rigel}   & $863$ ly  & $8.16\cdot(10^{-4}\div 10^{-2})$  \\ \hline
\mbox{Alnilam} & 2000 ly  & $1.89\cdot(10^{-3}\div 10^{-1})$  \\ \hline
\mbox{Deneb} & 2620 ly & $2.48\cdot(10^{-3}\div 10^{-1})$  \\ \hline
\mbox{WLM} & 3 Mly  & 2.83$\div$283  \\ \hline
\mbox{M31} & 145 Mly & $1.37\cdot(10^{2}\div 10^{4})$  \\ \hline
\mbox{NGC 2336} & 225 Mly & $2.12\cdot(10^{2}\div 10^{4})$  \\ \hline
\end{tabular}
\label{tableEins}
\end{center}
\end{table}
In the case a massive/massless particle reaches the screen for small values of the proper time/affine parameter (this can be achieved by positioning the screen close enough to the origin), we can use the linearizations
\begin{eqnarray}
\cos{{\left[kc(t_0+\sqrt{1+\gamma^2}\tau)+\varphi_{22}\right]}}-\cos{(kct_0+\varphi_{22})}&=&-kc\sqrt{1+\gamma^2}\sin{(kct_0+\varphi_{22})}\tau+\mathcal{O}(\tau^2),\\
\cos{{\left[kc(t_0+s)+\varphi_{22}\right]}}-\cos{(kct_0+\varphi_{22})}&=&-kc\sin{(kct_0+\varphi_{22})}s+\mathcal{O}(s^2)
\end{eqnarray}
in equations (\ref{ytau}) and (\ref{yms}) and if we further impose the conditions $y(\tau_s)=L$ and $y(\sigma_s)=L$, we find that a massive and a massless particle hit the screen at 
\begin{equation}\label{zeiten}
\tau_s=\frac{L}{c\gamma},\quad\sigma_s=\frac{L}{c}.
\end{equation}
Replacing (\ref{zeiten}) into (\ref{xtau}), (\ref{ztau}) and (\ref{xms}), (\ref{yms}) yields the following trajectories on the screen
\begin{eqnarray}
x(\tau_s)&=&0,\quad~~~~~~~~~~~~~~~~~~~~~~~~~~~~~~~~~~~~~~~~~~~~~~~~~~~~x(\sigma_s)=0,\\
z(\tau_s)&=&LH_{23}\left[\sin{(kct_0+\varphi_{23})}-\cos{(kct_0+\varphi_{23})}\right],\quad
z(\sigma_s)=z(\tau_s).\label{zts}
\end{eqnarray}
At this point a comment is in order. First of all, we observe that in this case the trajectory is the same both for massive and massless particles and it is represented by a vertical line segment on the $z$-axis. Moreover, in the case of an elliptically polarized gravitational wave with $\varphi_{23}=\varphi_{22}\pm\frac{\pi}{2}$, we have
\begin{equation}
z(\tau_s)=\pm LH_{23}\left[\sin{(kct_0+\varphi_{22})}+\cos{(kct_0+\varphi_{22})}\right].
\end{equation}
The same conclusions also hold for a massless particle. Finally, the amplitudes of these fluctuations are negligibly small because typical values of the strain $H_{\mu\nu}$ are of the order $10^{-22}$ so one would have from (\ref{zts}) a fluctuation of order smaller than $10^{-22}$ due to the fact that in the present scenario $L\ll 1$. What happens if we position the screen perpendicularly to the $z$-axis or the $x$-axis? In the first case, if we imagine to shoot a massive/massless particle in the $z$-direction with initial four-velocities given by  $U_0^\mu=(\sqrt{1+(U_0^3)^2},0,0,U_0^3)$ or $U_0^\mu=(U_0^3,0,0,U_0^3)$, respectively, and at the same time, we place the screen at $z=L\gg 1$, we find that a massive/massless particle hits the screen for
\begin{equation}
\widehat{\tau}_s=\frac{L}{c\beta}\left[1-h_{22}(0)\right],\quad
\widehat{\sigma}_s=\frac{L}{c}\left[1-h_{22}(0)\right], \quad\beta=\frac{U_0^3}{c}
\end{equation}
at the following point on the screen
\begin{eqnarray}
x(\widehat{\tau}_s)&=&-x(\tau_s),\quad x(\widehat{\sigma}_s)=-x(\sigma_s),\\
y(\widehat{\tau}_s)&=&z(\tau_s),\quad
y(\widehat{\sigma}_s)=z(\sigma_s),\\
z(\widehat{\tau}_s)&=&L,\quad
z(\widehat{\sigma}_s)=L,
\end{eqnarray}
where $x(\tau_s)$, $y(\tau_s)$, $x(\sigma_s)$ and $z(\sigma_s)$ are given by (\ref{mix1s}) and (\ref{mix3s}). The above result shows that the shape of the trajectories does not depend on whether the screen is positioned at $z=L$ or $y=L$. Finally, if we shoot a  massive particle along the $x$-axis with initial four-velocity $U_0^\mu=(\sqrt{1+(U_0^1)^2},U_0^1,0,0)$, equations (\ref{uu1})-(\ref{uu4}) become 
\begin{equation}
U^0=\sqrt{1+(U_0^1)^2},\quad
U^1=U_0^1,\quad
U^2=U^3=0.
\end{equation}
Integrating the above equations under the usual condition that $x(0)=y(0)=z(0)=0$ yields
\begin{equation}
t=t_0+\sqrt{1+(U_0^1)^2}\tau,\quad
x(\tau)=U_0^1\tau,\quad
y(\tau)=z(\tau)=0.
\end{equation}
From the above result we conclude that in the case of a massive particle moving in the same direction as the gravitational wave there will be no detectable effect on its trajectory. For massless particle we cannot use (\ref{uu1})-(\ref{uu4}) because the denominator $U^1_0-U^0_0$ vanishes in the present case. Hence, we need to go back to the original set of differential equations obtained by the Euler-Lagrange method. More precisely, if we integrate the set of equations
\begin{eqnarray}
\frac{d}{ds}\left[(1-h_{22})U^2-h_{23}U^3\right]&=&0,\\
\frac{d}{ds}\left[(1+h_{22})U^3-h_{23}U^2\right]&=&0,
\end{eqnarray}
we find that
\begin{eqnarray}
U^2&=k_1\left[1+h_{22}(s)\right]+k_2 h_{23}(s),\\
U^3&=k_2\left[1-h_{22}(s)\right]+k_1 h_{23}(s),
\end{eqnarray}
with integration constants
\begin{eqnarray}
k_1&=\left[1-h_{22}(0)\right]U_0^2-h_{23}(0)U_0^3,\\
k_2&=\left[1+h_{22}(0)\right]U_0^3-h_{23}(0)U_0^2.
\end{eqnarray}
Since $U_0^2=U_0^3=0$, we conclude that $k_1=k_2=0$ and hence, $U^2=U^3=0$. Moreover, $U^0$ and $U^1$ are governed by the equations
\begin{eqnarray}
\frac{dU^0}{ds}-\frac{1}{2}\frac{\partial h_{22}}{\partial t}\left[(U^2)^2-(U^3)^2\right]-\frac{\partial h_{23}}{\partial t}U^2 U^3&=&0,\\
\frac{dU^1}{ds}-\frac{1}{2}\frac{\partial h_{22}}{\partial t}\left[(U^2)^2-(U^3)^2\right]-\frac{\partial h_{23}}{\partial t}U^2 U^3&=&0,
\end{eqnarray}
which in view of the previous result they simplify as follows
\begin{equation}
\frac{dU^0}{ds}=0=\frac{dU^1}{ds}.
\end{equation}
Hence, also in this case the gravitational wave does not influence the
motion of the particle.

We conclude the present section by giving an interpretation of the
results from another perspective. First,
we compute the spatial velocity of a photon when it arrives at the
screen at the value $\sigma_s$ of the affine parameter (see
eq. (\ref{zeit})). Taking
into account that the spatial components of the velocity vector $\mathbf{V}=(V^1,V^2,V^3)$ are $V^i=U^i(ds/dt)$ and from (\ref{uu1}) in the limit $s\gg 1$
\begin{equation}
\frac{ds}{dt}=1+\frac{1}{2}h_{22}(0)-\frac{1}{2}h_{22}(s),
\end{equation} 
it follows that 
\begin{eqnarray}
V^1(\sigma_s)&=&-\frac{c}{2}h_{22}(0)+\frac{c}{2}\sin{(kL+\varphi_{22})},\quad
V^2(\sigma_s)= c\left[1-\frac{1}{2}h_{22}(0)\right]+\frac{c}{2}H_{22}\sin{(kL+\varphi_{22})},\\
V^3(\sigma_s)&=&-ch_{23}(0)+cH_{23}\sin{(kL+\varphi_{23})},\quad
|\mathbf{V}|=c\left(1-\frac{1}{2}h_{22}(0)+\frac{1}{2}H_{22}\sin{(kL+\varphi_{22})}\right),
 \end{eqnarray}
where w.l.o.g. we set $t_0=0$. The successive arrivals of the photons at
the screen will happen so fast that the photons impinging on the
screen will produce a ring of light (or part of a ring). It is then
instructive to imagine a cone defined by the point (apex) from which
the light ray due to gravitational deflection starts diverging (this
point can, of course, be the star itself) and a circle at the
distance $L_1$ from the apex with radius $R_1=L_1H$ describing the light
ring.
$L_1$ should be the distance at which the crest of the gravitational waves stop
arriving and, therefore, for $L>L_1$ the gravitational interaction on
the light is absent. Let $L_2$ be the distance from the apex to the
observation screen.  The ray at $L_1$ will continue its trajectory
in the direction of $\mathbf{V}$ with $L+L_1$ in the equation above.
Let $\textbf{R}=(x(\sigma_s),y(\sigma_s),z(\sigma_s))$ denote the spatial part of the position vector of the photon arriving at the screen at $s=\sigma_s$. Then, (\ref{xms})-(\ref{zms}) together with Werner's formula yields
\begin{eqnarray}
x(\sigma_s)&=&-\frac{L}{2}h_{22}(0)+\frac{H_{22}}{k}\sin{\left(\frac{kL}{2}+\varphi_{22}\right)}\sin{\left(\frac{kL}{2}\right)},\quad
y(\sigma_s)=L+2\frac{H_{22}}{k}\sin{\left(\frac{kL}{2}+\varphi_{22}\right)}\sin{\left(\frac{kL}{2}\right)},\\
z(\sigma_s)&=&-Lh_{23}(0)+\frac{2H_{23}}{k}\sin{\left(\frac{kL}{2}+\varphi_{23}\right)}\sin{\left(\frac{kL}{2}\right)},\quad
|\textbf{R}|=L+\frac{2}{k}\sin{\left(\frac{kL}{2}+\varphi_{22}\right)}\sin{\left(\frac{kL}{2}\right)}.
\end{eqnarray}
One can readily calculate the angle $\vartheta$ between $\mathbf{R}$
up to order $h$. Since $\mathbf{R}$ points to the ring at the screen, 
  the information on $\vartheta$ will give us an idea in which
  direction the light continues. After expansion we obtain
\begin{equation}
\cos{\vartheta}=\frac{\textbf{R}\cdot\textbf{V}}{|\textbf{R}||\textbf{V}|}=1+\mathcal{O}(h^2)
\end{equation}
which implies that the angle is almost zero. We do not give here the
$\mathcal{O}(h^2)$ corrections as they are not trustworthy using a
formalism valid only up to order $h$. In this order the light continues
along $\mathbf{R}$ magnifying the circle at a distance $L_2$ defining
a new cone with $L_2$ and the radius of the circle being $R_2$. Simple
intercept theorems tell us that
\begin{equation} \label{R2}
  R_2=R_1\frac{L_2}{L_1}=L_2H
\end{equation}
bearing in mind that $R_1=L_1H$. It is surprising that the result does
not depend on $L_1$ which we defined as the distance at which the
gravitational wave stops interfering with the light. In reality,
there will be such a dependence through terms of order $h^2$, but they
can only be calculated by extending the formalism from the beginning
up to $\mathcal{O}(h^2)$ (or using an exact solution). Here, we give an
estimate of the effect by using the results of the first order
formalism where at a distance $L_2$ we see a ring of the radius
$L_2H$.
Let us assume $L_2$ to be in the range of size of our galaxy, i.e.,
$L_2 \simeq 10^4$ pc. With $H=10^{-22}$ we obtain $R_2 \sim 1$ cm.
Hence, seeing the light ray across the galaxy exposed to a
gravitational wave will give an effect which could, in principle, be observable
(hereby not
taking into account all technological obstacles). If
it is possible to identify the light from a star situated in another
galaxy, the effect is even bigger.  A second order formalism should
eventually be taken into account to calculate more exactly the
trajectory of the light after having been exposed to the gravitational wave.

\section{Discussion}
We have re-calculated the effect of a gravitational wave on massive
and, in particular, massless particles (light ray). In some details, 
we defer from reference \cite{deFelice} where the topic has been
considered for the first time. We gave an interpretation in terms of a
cone defined by the star and the imaginary screen on which the photons
outline a ring of the radius $RH$ where $R$ is the distance between the
start and observer and $H$ the strain of the gravitational wave.  In
this interpretation, it is not necessary that the gravitational wave
impinges in the light ray up to the distance $R$. The main
result, valid at the first order in $h$,  is that the radius of the ring does not depend on the distance
where the gravitational effect ceases. We therefore think that the more
accurate picture of what is happening should be given by a second
order formalism or by a space-time matching procedure but the absence of free parameters makes this approach impossible. In spite of this, we think that the overall
description is correctly given by the first order formalism.
If so, an observation of this effect is not excluded as it is enhanced
with the distance $R$. We conclude this section with a couple of remarks. Rings of light appear also in the context of strong lensing where Einstein rings \cite{Falco} may form. However, the astrophysical object which cause the Einstein ring has a larger lifespan than the one producing the ring of light discussed here. The difference is in the time window we can see such a ring since the typical characteristic time to see the effect will be much shorter. Concerning the instabilities of the source itself, we are aware that in the case of neutron stars the presence of $p$-, $f$- and $r$-gravitational waves \cite{inst} whose decay in time may vary considerably can affect the ring brightness. This kind of waves have not been implemented in our model and this would be an interesting aspect to be considered in future work. Moreover, in our model we assumed that the observer on the screen does not move nor is affected by the transverse gravitational waves. We will leave the analysis of the corresponding problem where the aforementioned assumptions are relaxed as an interesting topic for a future paper.

\bibliography{sample}

\section*{Acknowledgements}
We are grateful to the anonymous referees for their comments that helped to improve the present work.
\section*{Author contributions statement}
M.N. conceived the problem discussed in this work and contributed to the interpretation of the results. N.M.B. participated in the preparation of section 2.   D.B. and J.M.F. worked out all of the technical details, performed the  calculations. D.B. wrote the manuscript in consultation with M.N.. All authors reviewed the manuscript. 
\section*{Additional information}
Davide Batic declares that there are no conflicts of interest for all authors in the present manuscript.
\section*{Availability of Data and Materials}
All data generated or analysed during this study are included in this published article.
\end{document}